\documentclass[%
 reprint,
 amsmath,amssymb,
 aps,
]{revtex4-2}

\usepackage{graphicx}% Include figure files
\usepackage{dcolumn}% Align table columns on decimal point
\usepackage{bm}% bold math
\begin{document}

\preprint{APS/123-QED}

\title{Propagation effects in resonant high-harmonic generation \\ and high-order frequency mixing in laser plasma}

\author{Vasily V. Strelkov}
 \email{strelkov.v@gmail.com}
 \affiliation{Prokhorov General Physics Institute of the Russian Academy of Sciences, Vavilova Street 38, Moscow 119991, Russia}
\author{Margarita A. Khokhlova}%
 \affiliation{%
 King’s College London, Strand Campus, London WC2R 2LS, UK}

\begin{abstract}
We study the phase matching of resonant high-harmonic generation~(HHG) and high-order frequency mixing~(HFM) in plasma. We solve numerically the propagation equations coupled with the time-dependent Schr\"odinger equation for the nonlinear polarisation. The macroscopic harmonic signal is enhanced in the vicinity of a multiphoton resonance with the transition between the ground and autoionising states of the generating ion. We show that narrow and strong resonances (as for gallium and indium ions) provide compensation of the plasma dispersion in a spectral region above the exact resonance improving the phase matching and leading to a high macroscopic signal. The compensation does not take place for a wider resonance (as for manganese ions), instead the phase matching is achieved in the HFM process. Comparing the XUV generated in manganese plasma and in neon gas, we show that the resonant HHG in plasma is an order of magnitude more effective than in the gas, moreover another order of magnitude can be gained from the propagation using HFM in plasma.
\end{abstract}

\maketitle

Attosecond pulses have proven to be a unique instrument to observe and control ultrafast electron dynamics in different kinds of matter~\cite{Villeneuve2018, Ryabikin2023, Khokhlova2023}, and the improvement of attopulse brightness is of key fundamental and technological importance. 

The production of attopulses largely relies on the process of high-harmonic generation~(HHG), where many photons of the strong laser driver are converted into a comb of high harmonics~(HH) through a highly nonlinear nonperturbative response of the medium. The total efficiency of this process is composed of the microscopic, single-particle, response and of the macroscopic response describing the propagation of the light through the medium. Commonly, one of these responses is improved to reach the higher efficiency of HHG, and thus brighter attopulses: microscopic, by choosing the generating medium smartly, or macroscopic, by optimising the propagation conditions such as the density of the medium and the laser parameters.
%\red{: intensity, wavelength, pulse duration, beam profile and so on}.

One way to boost the microscopic HHG response is to use a resonance of HHs and a transition between the states of the generating system. In the case of the transition between the ground state and an autoionising state~(AIS), the resonant HHs are generated up to orders of magnitude more efficiently than the nonresonant ones~\cite{Singh2021}. Resonant HHG enhancement has been observed in plasma plumes~\cite{Singh2024, Ganeev2016, Fu2022, Fu2023, Singh2021}, as well as for a `giant' resonance in Xe~\cite{Shiner2011}. 
%\red{It has also been shown that HHG using \SI{400}{nm} in plasma plumes~\cite{Singh2021, Fu2023} achieves a much higher efficiency than using \SI{800}{nm} driver}. 

For the HHG propagation, it has been shown~\cite{Constant, KhokhlovaStrelkov} that there are three mechanisms limiting the coherent growth of the HH energy with propagation length: absorption, phase matching and blue shift. However, there is other fundamentally-close high-order process, taking place when the strong driver is accompanied by another generating field~--- high-order frequency mixing~(HFM)~\cite{Eichmann1995, Gaarde1996, Balcou1999}~--- which  experiences the phase-matching and blue-shift limitations to lesser extent~\cite{Shkolnikov, Milchberg, Chichkov_plasma, Platonenko_non-collinear, KhokhlovaStrelkov, Kapteyn_2007, Hort2021}.

%In HFM the generation of an XUV photon involves several photons from the fundamental and several photons from the additional field. Thus, in the microscopic response, on top of HHs, there are new frequency components corresponding to a different number of quanta from the generating fields. It has already been shown in early studies~\cite{Shkolnikov, Milchberg, Chichkov_plasma, Shkolnikov_1996, Platonenko_non-collinear, Chichkov_2000} that the phase matching improves for HFM, in particular, for the phase-matched generation of specific HFM components of relatively low order in plasma~\cite{Shkolnikov, Milchberg, Chichkov_plasma}, followed by further studies in this direction~\cite{Kapteyn_2007, Worner_non-collinear, PhysRevLett.112.143902, Oguchi_PRA, Strelkov_np, Ellis_2017, Tran_2019, Harkema:19, Ganeev_2016, KhokhlovaStrelkov, Birulia2024}.

In this Letter we study resonant HHG, as a process resulting in a boosted microscopic response, considering theoretically the macroscopic properties of this process. We also investigate the resonant HFM, optimizing the propagation by choosing parameters of the additional weak generating field.

We start from the analysis of the propagation properties of the HHG process along with the microscopic resonant response. For HHG the coherence length is defined as
\begin{equation}
L_\mathrm{coh} =\frac{\pi}{|\Delta k |} \, ,
\label{L_coh}
\end{equation}
where the wavevector mismatch is $\Delta k =k_q - q k_0$, here $q$ is a HH order, $k_0$ and $k_q$ are the driver and HH wavevectors, correspondingly. The wavevectors can be written as 
\begin{equation}
k_0 =\frac{\omega_0}{c}\left(1+\Delta n_0^\mathrm{(ion)}+\Delta n_0^\mathrm{(el)}\right) + \delta k_0^\mathrm{(geom)} \, ,
\label{k_0}
\end{equation}
\begin{equation}
k_q =\frac{q \omega_0}{c}\left(1+\Delta n_q^\mathrm{(ion)}+\Delta n_q^\mathrm{(el)}\right) \, ,
\label{k_q}
\end{equation}
where $\omega_0$ is the driver frequency, $\Delta n_0^\mathrm{(el)}$ is the contribution to the refractive index of the driver due to the free electrons written (atomic units are used throughout if not specified otherwise) as 
\begin{equation}
\Delta n_0^\mathrm{(el)}=-\frac{2 \pi N }{ \omega_0^2} \, ,
\label{Dn_0}
\end{equation}
with the electronic density $N$ (in usual units the latter equation is written as $\Delta n_0^\mathrm{(el)}=-4.5 \times 10^{-22} N[$cm$^{-3}](\lambda_0[\mu$m$])^2$); $\Delta n_q^\mathrm{(ion)}$ is the additive to the refractive index for the harmonic due to the ions. Note that the common assumption for the HHG phase matching is that the HH refractive index is unity. While it is a very reliable assumption for nonresonant HHG, here we show that this is not the case for the resonant one. Namely, the resonant term for $\Delta n_q^\mathrm{(ion)}$ is written as
\begin{equation}
\Delta n_q^\mathrm{(ion)}=-\frac{\pi N  f}{ \Omega} \quad \frac{\Delta \omega}{\Delta \omega^2+ \Gamma^2/4} \, ,
\label{Dn_q}
\end{equation}
where $\Omega$ and $\Gamma$ are the resonant frequency and width, $\Delta \omega=q \omega_0-\Omega$  is the detuning from the resonance, $f$ is the oscillator strength of the resonant transition, and we assume that the ionic and the electronic densities are equal. 

The other terms contributing to the medium refractive index in Eqs.~(\ref{k_0}) and~(\ref{k_q}) are negligible: $\Delta n_q^\mathrm{(el)}$ is rather small because the HH frequency is high, while $\Delta n_0^\mathrm{(ion)}$ is low because the fundamental frequency is far from ionic resonances. We also neglect the geometric dispersion $\delta k_0^\mathrm{(geom)}$, which can be taken into account similarly to the $\Delta n_0^\mathrm{(el)}$ contribution if necessary.

One can see from Eqs.~(\ref{Dn_0}) and (\ref{Dn_q}) that above the resonance the contribution of the ions to the detuning can compensate the contribution of the free electrons. The maximum value of $|\Delta n_q^\mathrm{(ion)}|$ is achieved near the resonance for $|\Delta \omega|= \Gamma/2$. For $\Delta \omega= \Gamma/2$:
\begin{equation}
%\frac{\Delta n_q^\mathrm{(ion,max)}}{\Delta n_0^\mathrm{(el)}}=\frac{f \omega_0^2 }{2 \Gamma \Omega} \, .
\Delta n_q^\mathrm{(ion,max)} / \Delta n_0^\mathrm{(el)} = f \omega_0^2 / (2 \Gamma \Omega) \, .
\label{ratio}
\end{equation}

The absorption of the generated radiation near the resonance is also important. The absorption length is $L_\mathrm{abs}=1/(N \sigma)$ with the photoionisation cross-section $\sigma$, which can be written using the parameters of the resonance (see, for instance,~\cite{Singh2021}) as
\begin{equation}
L_\mathrm{abs}=\left[ \frac{\pi N f}{c} \quad \frac{\Gamma}{\Delta \omega^2+\Gamma^2/4} \right]^{-1} \, .
\label{L_abs}
\end{equation}
From Eqs.~(\ref{L_abs}) and (\ref{L_coh})-(\ref{Dn_q}) for $\Delta \omega \ll \Omega$, we can write
\begin{equation}
\frac{L_\mathrm{abs}}{L_\mathrm{coh}}=\left| \frac{2 \Omega (\Delta \omega^2 + \Gamma^2/4)}{\pi \Gamma \omega_0^2 f }-\frac{\Delta \omega }{\pi \Gamma}\right| \, .
\label{L_abs_L_coh_0}
\end{equation}
In particular, near the resonance ($\Delta \omega = \Gamma/2$) we have
\begin{equation}
\frac{L_\mathrm{abs}}{L_\mathrm{coh}} 
= \frac{1}{2 \pi} \left| 1- \frac{2 \Gamma \Omega}{f \omega_0^2}\right| 
= \frac{1}{2 \pi} \left| 1- \frac{\Delta n_0^\mathrm{(el)}}{\Delta n_q^\mathrm{(ion,max)}}\right| \, .
\label{L_abs_L_coh}
\end{equation}
% or (see Eq.~(\ref{ratio}))
% $$
% \frac{L_\mathrm{abs}}{L_\mathrm{coh}}=\frac{1}{2 \pi} \left| 1- \frac{\Delta n_0^\mathrm{(el)}}{\Delta n_q^\mathrm{(ion,max)}}\right| \, .
% $$

From the latter equation we see that when the electronic contribution to the dispersion dominates, HHG is phase-matching limited ($L_\mathrm{coh} \ll L_\mathrm{abs}$); this is the case for high $\Gamma$, $\Omega$ and low $f$, $\omega_0$. For the opposite case, free-electron and ion contributions can compensate each other (such compensation was discussed in~\cite{GaneevPRA2013}), so the macroscopic HHG is absorption-limited.

We study several resonant HHG conditions. We consider the transition  $3d^{10} 4s^2 \quad ^1 S_0 \xrightarrow{} 3d^9 4s^2 4p \quad  ^1 P_1$ in GaII (the transition energy is~21.88~eV, see~\cite{Pindzola1982, Peart1987, Singh2021} for other parameters), the transition   $4d^{10} 5s^2 \quad  ^1 S_0  \xrightarrow{} 4d^9 5s^2 5p (^2D)\quad  ^1 P_1$ in InII (the transition energy is~19.92~eV, see~\cite{Duffy2001} for other parameters) and the `giant' $3p \xrightarrow{} 3d$ resonance in MnII (the transition energy is~51~eV, see~\cite{Kjeldsen2004} for other parameters).

We find that for the IR driver for all three cases the ratio~(\ref{ratio}) is below unity, thus HHG is phase-matching limited. However, when we switch to a visible driver, the situation becomes complicated. For the resonant H7 generation in GaII (driver wavelength is $\sim$~400~nm%
\footnote{%
It has been shown that HHG with~400~nm driver in plasma plumes~\cite{Singh2021, Fu2023} achieves a much higher efficiency than using an~800~nm one.%
}%
, which is the second harmonic of the Ti:Sapph laser), we find that ratio~(\ref{ratio}) is 0.96, and for the resonant H7 generation in InII (driver wavelength is $\sim$~435~nm), this ratio is as high as 2.1. Thus, above the exact resonance the contribution of the free electrons can be compensated by the ionic contribution in Ga as well as in In plasma. In the vicinity of the frequencies, for which the compensation takes place, the HHG is absorption-limited, while outside this bandgap it is phase-matching limited, see Eq.~(\ref{L_abs_L_coh_0}). For the resonant H17 generation in Mn plasma (driver wavelength is $\sim$~420~nm) the ratio~(\ref{ratio}) is 0.034, thus the HHG is phase-matching limited even using a visible generating field. 

The wavevector mismatch, which is a key factor limiting the HHG macroscopic efficiency in this case, can be essentially reduced for the process of HFM. 
%This process occurs when a second generating field is added to a strong fundamental one~\cite{Eichmann1995, Gaarde1996, Balcou1999}. 
%In HFM the generation of an XUV photon involves several photons from the fundamental and several photons from the additional field. Thus, in the microscopic response, on top of HHs, there are new frequency components corresponding to a different number of quanta from the generating fields. It has already been shown in early studies~\cite{Shkolnikov, Milchberg, Chichkov_plasma, Shkolnikov_1996, Platonenko_non-collinear, Chichkov_2000} that the phase matching improves for HFM, in particular, for the phase-matched generation of specific HFM components of relatively low order in plasma~\cite{Shkolnikov, Milchberg, Chichkov_plasma}, followed by further studies in this direction~\cite{Kapteyn_2007, Worner_non-collinear, PhysRevLett.112.143902, Oguchi_PRA, Strelkov_np, Ellis_2017, Tran_2019, Harkema:19, Ganeev_2016, KhokhlovaStrelkov, Birulia2024}.
Let us consider the HFM process involving $Q$ quanta from the strong fundamental and $m$ quanta from the additional weak field. The HFM process leads to generation of frequency components
\begin{equation}
\omega_{Q,m}=Q \omega_0 +m \omega_1 \, ,
\label{HFMfreqs}
\end{equation}
where $Q$, $m$ and $Q+m$ are natural, integer and odd numbers, respectively. For a weak additional field in comparison to the driver, HFM leads to a notable efficiency only for processes involving few weak-field quanta: $|m| \ll Q$. If the plasma contribution dominates in the dispersion, the phase mismatch neglects~\cite{Shkolnikov, Milchberg, Chichkov_plasma, Platonenko_non-collinear, KhokhlovaStrelkov, Kapteyn_2007} for this process for $m<0$ and for the frequency of the weak field
\begin{equation}
\omega_1= \frac{|m|}{Q} \omega_0 \, , 
\label{Q}
\end{equation}
which means that the weak-field frequency is much lower than the one of the driver. Note that for orders $q$ that neighbour the ideally phase-matched $Q$, the phase mismatch is rather small.

One more phenomenon that influences the macroscopic efficiency of HHG and HFM is the plasma-induced blue shift. The ionisation of the medium by the generating fields leads to frequency shifts of these fields. In spite of being small in comparison to the fields’ frequencies, these shifts are multiplied by a high order of the process that leads to a noticeable shift of the generated frequencies. The latter shift can limit the macroscopic efficiency of the HHG and HFM processes~\cite{KhokhlovaStrelkov, Birulia2024}. It was shown~\cite{Hort2021} that for HFM the compensation of the plasma-induced blue shifts is similar to the compensation of the phase mismatch. Thus, the HFM process described by \eqref{HFMfreqs} and \eqref{Q} for $q \approx Q$ almost does not suffer from the blue shift of the generating fields.

Below we study numerically the HHG and HFM macroscopic signals taking into account the propagation of both the generating and generated fields in Ga and Mn plasma. We assume that initially the plasma consists of singly-charged ions and free electrons. The 1D propagation equations for the fields are integrated as described in~\cite{Birulia2024, Strelkov2023_photonics} using single-ion response calculated via numerical solution of the 3D time-dependent Schr\"odinger equation~(TDSE) at every propagation step. The TDSE is solved for the ionic model potential (similar to the one used in~\cite{Strelkov2010, Singh2021}) with the parameters chosen to reproduce the properties of the real ion: the ionisation energy, the AIS energy and its width. 

To understand the propagation of the XUV field in plasma, we numerically calculate the refractive index and the absorption coefficient in absence of the laser field, that defines the actual HH refraction and absorption only approximately because the laser field modifies them.

\begin{figure}
\centering
\fbox{\includegraphics[width=0.85\linewidth]{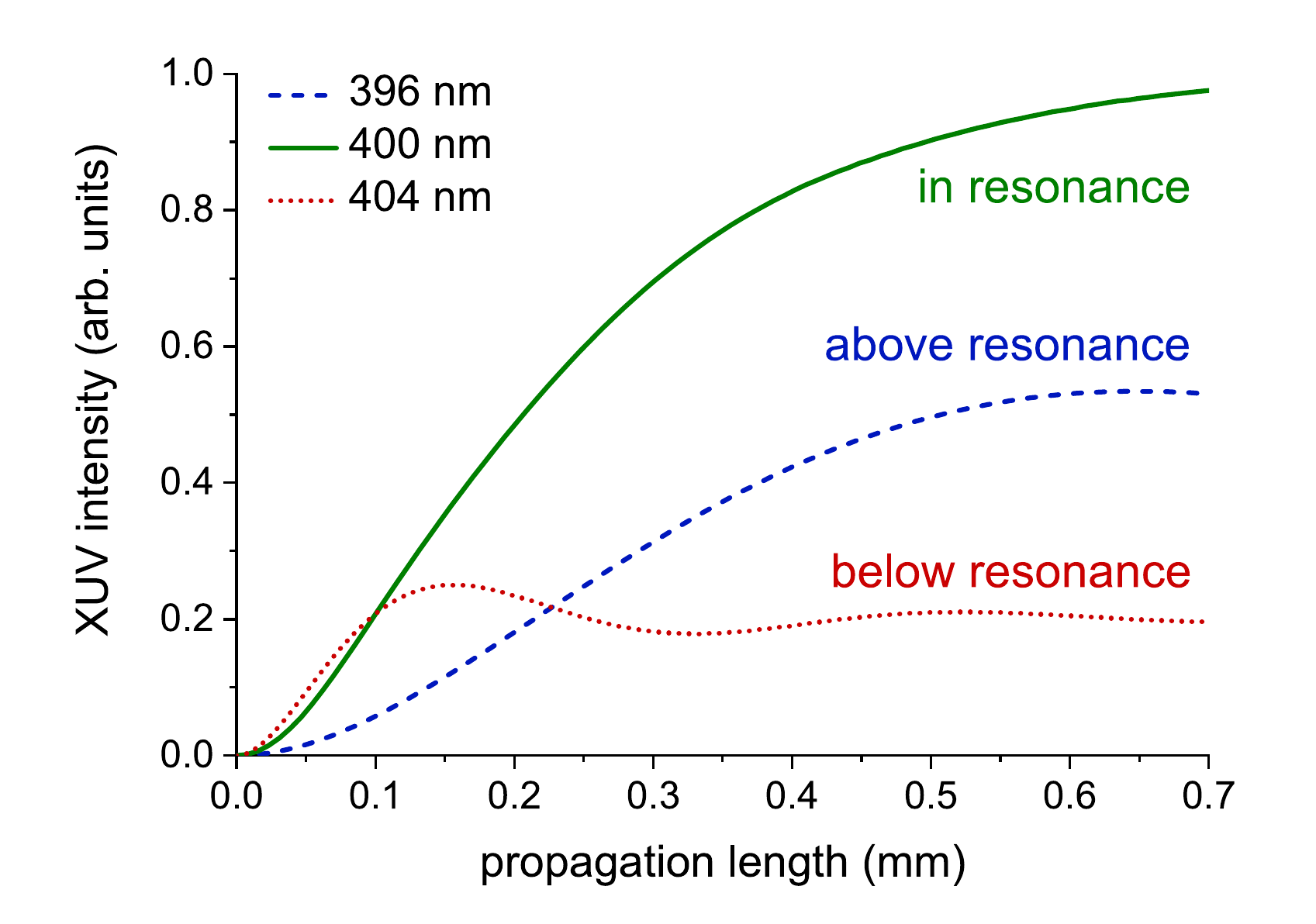}}
\caption{H7 macroscopic signals generated in Ga plasma as a function of propagation length. The drivers' wavelengths are 396~nm (above resonance), 400~nm (in resonance), 404~nm (below resonance). The laser pulse intensity is $3 \times 10^{14}$~W/cm$^2$, its duration is 60~fs, and the plasma density is $10^{18}$cm$^{-3}$.}
\label{Ga_int_vs_length}
\end{figure}

First, we study the effect of the detuning from the exact resonance between the harmonic energy and the ground-AIS transition energy on the propagation of the resonant HHG emission. Fig.~\ref{Ga_int_vs_length} shows the calculated H7 intensity in Ga plasma for three driving wavelengths in the vicinity of the 7-photon resonance with the ground-AIS transition. When the driving wavelength is such that H7 is in the exact resonance, as well as when H7 is above the exact resonance, the harmonic signal monotonically saturates with the propagation length. This means that the generation is absorption-limited in agreement with our analytical model above. When the fundamental frequency is such that `resonant' H7 is below the resonance, the initial growth of the macroscopic signal is stronger, but it stops growing at a much shorter distance. This behaviour is followed by oscillations of the harmonic signal, which is typical for phase-matching limited generation. Thus, even a minor variation of the driving wavelength (resulting in the detuning of H7 near the resonance) changes the behaviour of its macroscopic signal.
\begin{figure}[h!]
\centering
\fbox{
\includegraphics[width=0.9\linewidth]{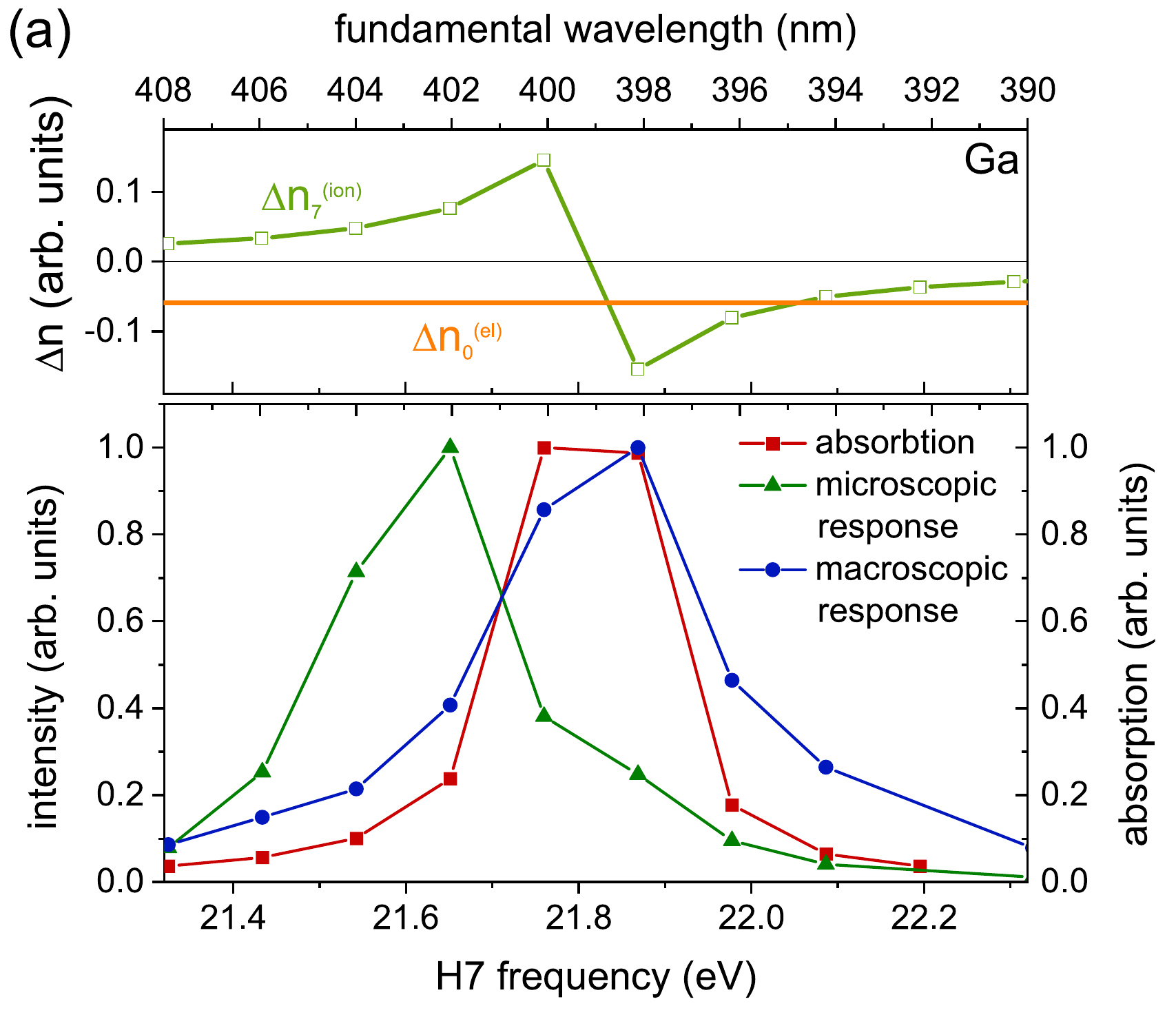}
}
\fbox{
\includegraphics[width=0.9\linewidth]{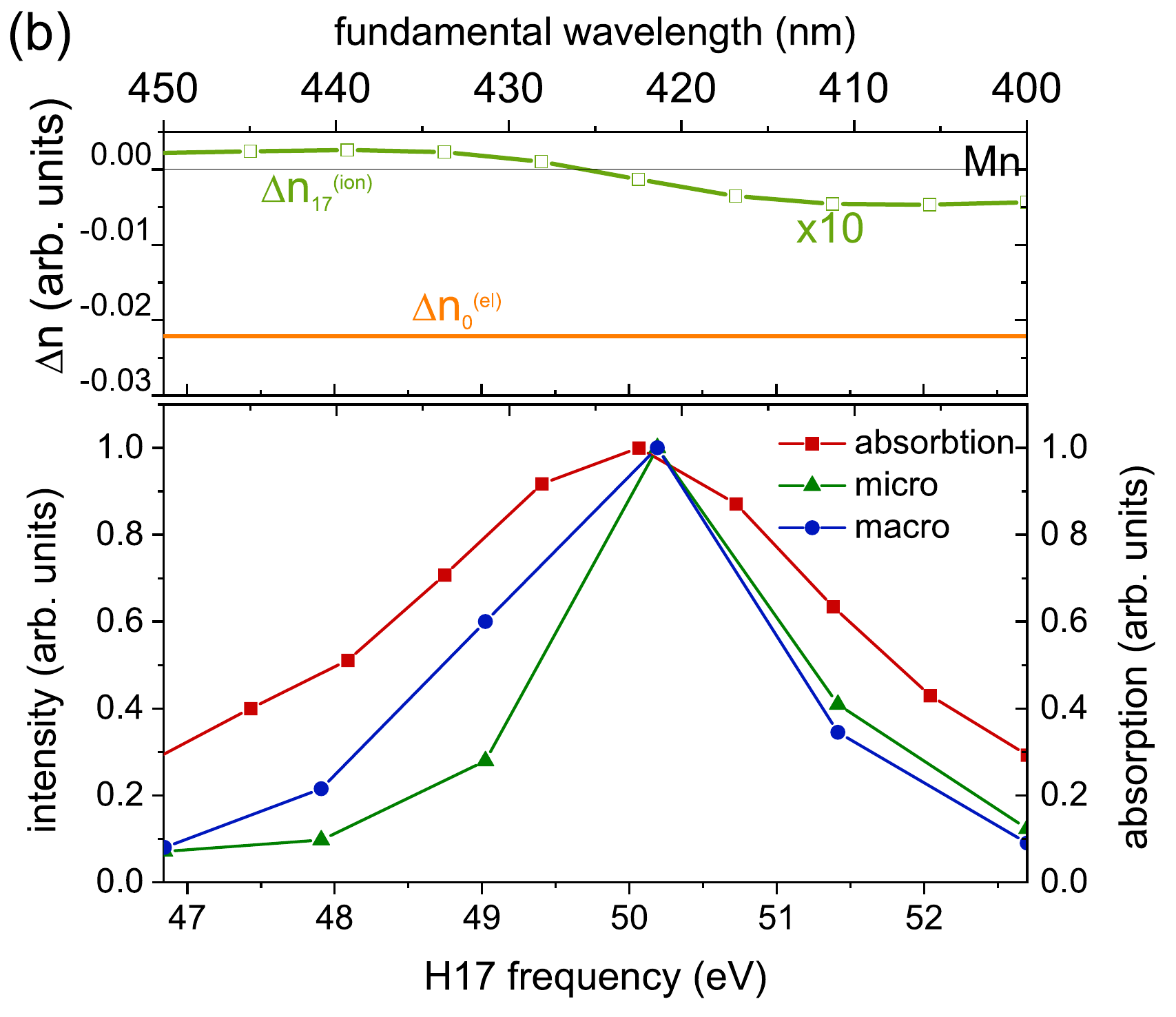}
}
\caption{XUV absorption line and intensities of microscopic and macroscopic (a)~H7 responses of Ga and (b)~H17 responses of Mn plasma for different fundamental wavelengths. Other laser pulse parameters for (a) are the same as in Fig.~\ref{Ga_int_vs_length}, and for (b) the driving pulse intensity is $9 \times 10^{14}$~W/cm$^2$, its duration is 40~fs. The upper panels at each graph show two main additives to the refractive indexes of the driver and of the resonant HH, defining phase matching of the HHG.}
%, see text for more details.}
\label{Ga_int_vs_freq}
\end{figure}

To understand this feature in more detail, we perform the calculations for a number of driving wavelengths, see the results in Fig.~\ref{Ga_int_vs_freq}. The macroscopic signal presented in the figure is the maximal HH macroscopic intensity achieved during propagation in the target. We see that the absorption line and the resonant enhancement of the microscopic HHG response do not coincide. For Ga plasma (Fig.~\ref{Ga_int_vs_freq}a), at first glance, the generation conditions {\it below} the absorption line are favourable due to the high microscopic response and low absorption. However, Fig.~\ref{Ga_int_vs_length} as well as the macroscopic response in Fig.~\ref{Ga_int_vs_freq}a show higher signal {\it above} the resonance.

To clarify this, Fig.~\ref{Ga_int_vs_freq} shows also the numerically calculated two main contributions to the refractive indexes, which define the wavevector mismatch for HHG in plasma, see Eqs.~(\ref{k_0}) and~(\ref{k_q}).
%The fundamental refractive index detuning from unity is mainly due to free electrons, and the XUV refractive index detuning is mainly due to the resonant contribution of the ions.
One can see that  above the resonance this two terms compensate each other improving phase matching for HHG. 
%The possibility of such compensation was discussed in~\cite{GaneevPRA2013}.
As discussed above, such compensation is possible for strong and narrow resonances (such as ones in GaII and InII), and only using visible or UV fundamental for which the contribution $\Delta n_0^{(el)}$ is moderate. 
%This contribution is proportional to the wavelength squared, so for IR fundamental it is much higher and the compensation is hardly possible  (for instance, using IR with a wavelength of approximately 800 nm one has the deviation of $n_0$ from unity four times higher than what is shown in the Fig.~\ref{Ga_int_vs_freq}). Moreover, in the inset one can see that the phase-matching below the resonance is weak because both resonant and plasma detunings have the same sign.

When H7 in Ga plasma is below the resonance, the microscopic response is high, but the phase matching is weak and HHG is limited by both absorption and phase matching, providing the weakest signal shown in Fig.~\ref{Ga_int_vs_length}. Above the resonance the microscopic response is weaker but the phase matching is better, providing higher HHG signal. The XUV absorption plays a key role in the saturation of this signal, as well as for the resonant case.

For the nonresonant harmonics, the plasma contribution to dispersion dominates, leading to a poor phase matching for their generation. Thus, the resonant harmonic is enhanced with respect to the neighboring ones not only due to microscopic, but also due to macroscopic response properties. For the resonant HHG in Ga plasma~\cite{Singh2021}  the observed enhancement of the macroscopic signal was higher than the calculated in the same paper enhancement of the microscopic signal.  This can be attributed to the described above favorable phase matching of the resonant HHG.

%\purple{For the nonresonant harmonics, the plasma contribution to dispersion dominates, leading to a poor phase matching for their generation. Thus, the resonant harmonic is enhanced with respect to the neighbouring ones not only due to microscopic, but also due to macroscopic response properties. This was illustrated for the resonant HHG in Ga plasma~\cite{Singh2021}, where the observed enhancement of the macroscopic signal is higher than the calculated enhancement of the microscopic signal, thereby agreeing with the described above favourable phase matching of the resonant HHG.}

% \purple{For the nonresonant harmonics, the plasma contribution to dispersion dominates leading to poor phase matching for their generation. Thus, the resonant harmonic is enhanced with respect to the neighboring ones not only due to microscopic, but also due to macroscopic response properties. Note that for the resonant HHG in Ga plasma~\cite{Singh2021} the observed enhancement of the macroscopic signal was higher than calculated in the same paper enhancement of the microscopic one. This can be attributed to the described above better phase-matching of the resonant HHG.}

\begin{figure}
\centering
\fbox{\includegraphics[width=0.9\linewidth]{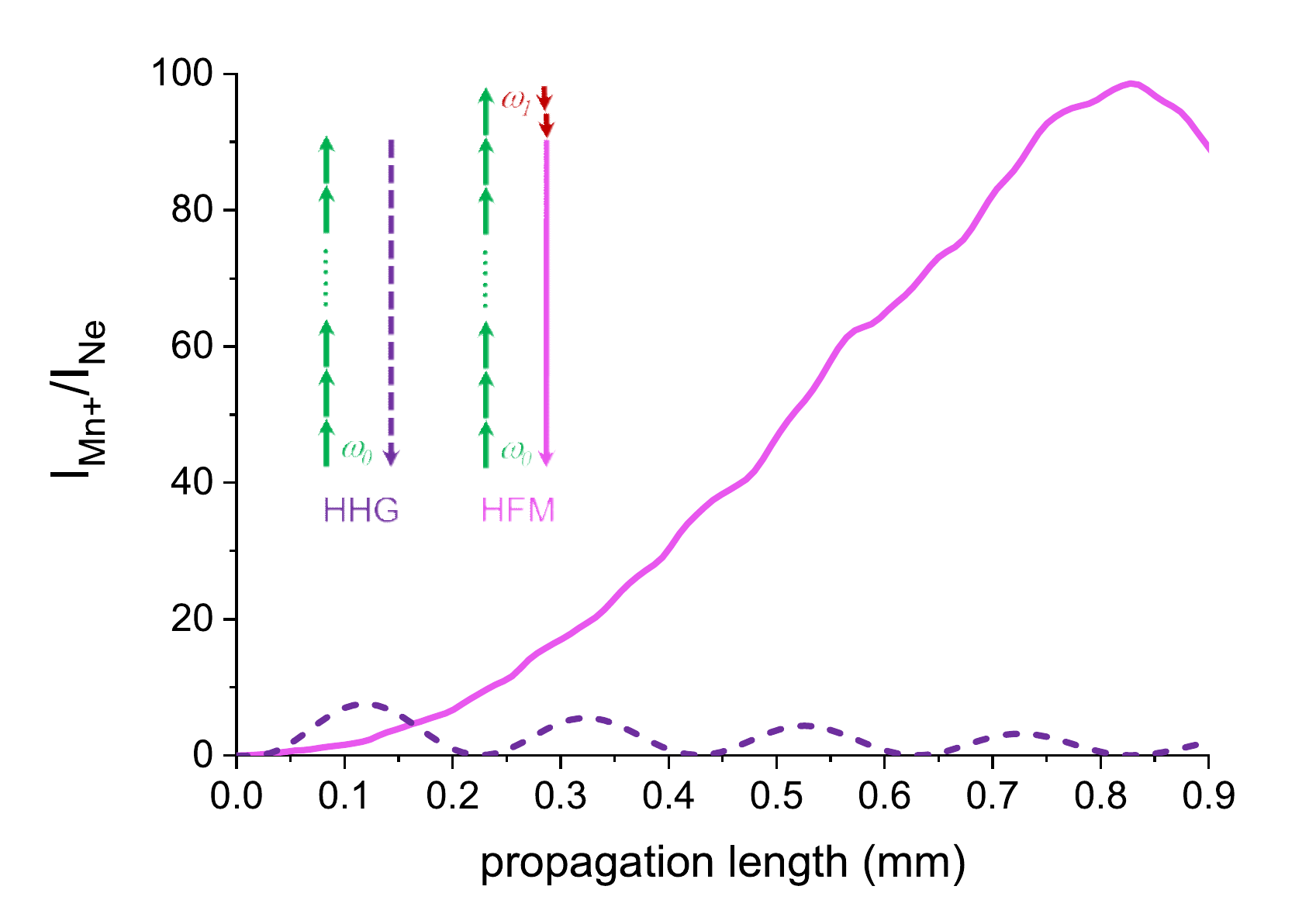}}
\caption{Macroscopic HHG and HFM signals in Mn plasma with propagation. The XUV intensities are divided by the maximal XUV intensity from HHG in Ne. The driver wavelength for HHG is 420~nm and for HFM is 415~nm, the weak-field wavelength is 3800~nm. The driving pulse intensity and duration are as in Fig.~\ref{Ga_int_vs_freq}b, the weak-field intensity is 0.5\% of the driver. The inset shows the sketch of the HHG and HFM processes.}
\label{Mn_int_vs_length}
\end{figure}

We also study HHG in Mn plasma in the vicinity of the resonance. Fig.~\ref{Ga_int_vs_freq}b presents the absorption coefficient as well as the microscopic and macroscopic responses. %calculated in Mn plasma.
One can see that the peaks of the three lines coincide, but their widths are different. The figure also shows the main contributions to the refraction coefficients for H17 and for the driver.
We see that the phase-matching conditions qualitatively differ from the ones in Ga plasma: in Mn plasma the resonant refraction coefficient is too low to compensate the plasma dispersion (in agreement with our analytical model), so HHG is phase-matching limited in the resonant case, as well as below and above it. Fig.~\ref{Mn_int_vs_length} shows the resonant H17 intensity with propagation length. One can see deep oscillations of the macroscopic signal, which are typical for phase-matching-limited generation.    
%
% \begin{figure}%[ht]
% \centering
% \fbox{\includegraphics[width=\linewidth]{Fig3.JPG}}
% \caption{Same as Fig.~\ref{Ga_int_vs_freq} for H17 generated in Mn plasma. The driving pulse intensity is $\SI{9e14}{W/cm^2}$, its duration is \SI{40}{fs}.}
% \label{Mn_int_vs_freq}
% \end{figure}

To characterise quantitatively the HHG efficiency in plasma, we calculate the HHG macroscopic signal in (initially neutral) Ne. Note that a rather high driving intensity is necessary to achieve a cutoff higher than H17 of the 400~nm fundamental. We choose Ne because HHG in Ne is compatible with such a high fundamental intensity. In Fig.~\ref{Mn_int_vs_length} one can see that the H17 generation efficiency in Mn plasma can be an order of magnitude higher than in Ne. This is achieved due to the very high resonant microscopic response of MnII and in spite of the weak phase matching of HHG in Mn plasma.

As we discussed above, the efficiency of phase-matching-limited HHG can be improved by realising HFM. The phase mismatch vanishes for harmonics with orders $q \approx Q$, see Eq.~(\ref{Q}). The highest conversion efficiency is achieved for $q$ slightly lower than $Q$~\cite{Birulia2024}; so to optimise HFM with $q=17$, we choose $Q=19$ and $m=-2$ and from Eq.~(\ref{Q}) we find that the low-frequency field wavelength is 3800~nm. In Fig.~\ref{Mn_int_vs_length} we see that the improved phase matching for HFM allows a growth of the signal up to a much longer propagation length than for HHG. This leads to a higher macroscopic response which exceeds the one for HHG in Ne by two orders of magnitude.

To summarise, we study theoretically the phase matching of resonant HHG and HFM in plasma. We integrate the propagation equations using the nonlinear polarisation found via numerical solution of the TDSE at every propagation step. The macroscopic HH signal is enhanced in the vicinity of multiphoton resonances with the transition between the ground state and AIS of the generating ion. We show that narrow and strong resonances in Ga~II (same as in In~II) ions provide compensation of the plasma dispersion in the region above the exact resonance. As a result, the improved phase matching leads to a high macroscopic signal in this spectral region. However, the compensation does not take place for wider resonances, as in Mn plasma, where the phase matching can be achieved instead in the HFM process. We compare the HHG efficiency in Ne gas and Mn plasma and show that the resonant HHG in plasma is an order of magnitude more effective than in the gas, while another order of magnitude can be gained switching from HHG to HFM in plasma.  

% WE NEED TO CUT REFS

Funding: This study was funded by RSF (Grant No. 24-22-00108).

Acknowledgments: MK acknowledges Royal Society funding under URF\textbackslash R1\textbackslash 231460.

\bibliography{lit}% Produces the bibliography via BibTeX.

\end{document}